\documentclass[reprint,aip,apl,amsmath,amssymb]{revtex4-2}

\usepackage[dvipdfmx]{graphicx}
\usepackage{dcolumn}
\usepackage{bm}

\usepackage[utf8]{inputenc}
\usepackage[T1]{fontenc}
\usepackage{mathptmx}

\usepackage{color}

\begin{document}

\preprint{AIP/123-QED}

\title{Ultraviolet photon-counting single-pixel imaging}
\author{Jun-Tian Ye}
 \altaffiliation{These authors contributed equally to this work.}
\author{Chao Yu}%
 \altaffiliation{These authors contributed equally to this work.}
\affiliation{Hefei National Research Center for Physical Sciences at the Microscale and School of Physical Sciences, University of Science and Technology of China, Hefei 230026, China}
\affiliation{Shanghai Research Center for Quantum Science and CAS Center for Excellence in Quantum Information and Quantum Physics, University of Science and Technology of China, Shanghai 201315, China}

\author{Wenwen Li}%
\author{Zheng-Ping Li}%
\affiliation{Hefei National Research Center for Physical Sciences at the Microscale and School of Physical Sciences, University of Science and Technology of China, Hefei 230026, China}
\affiliation{Shanghai Research Center for Quantum Science and CAS Center for Excellence in Quantum Information and Quantum Physics, University of Science and Technology of China, Shanghai 201315, China}
\affiliation{Hefei National Laboratory, University of Science and Technology of China, Hefei 230088, China}

\author{Hai Lu}
\author{Rong Zhang}
\affiliation{School of Electronic Science and Engineering, Nanjing University, Nanjing 210023, China}

\author{Jun Zhang}
 \altaffiliation{Authors to whom correspondence should be addressed: [Jun Zhang, zhangjun@ustc.edu.cn; Feihu Xu, feihuxu@ustc.edu.cn]}
\author{Feihu Xu}
 \altaffiliation{Authors to whom correspondence should be addressed: [Jun Zhang, zhangjun@ustc.edu.cn; Feihu Xu, feihuxu@ustc.edu.cn]}
\author{Jian-Wei Pan}
\affiliation{Hefei National Research Center for Physical Sciences at the Microscale and School of Physical Sciences, University of Science and Technology of China, Hefei 230026, China}
\affiliation{Shanghai Research Center for Quantum Science and CAS Center for Excellence in Quantum Information and Quantum Physics, University of Science and Technology of China, Shanghai 201315, China}
\affiliation{Hefei National Laboratory, University of Science and Technology of China, Hefei 230088, China}

\date{\today}

\begin{abstract}
We demonstrate photon-counting single-pixel imaging in the ultraviolet region. Toward this target, we develop a high-performance compact single-photon detector based on a 4H-SiC single-photon avalanche diode (SPAD), where a tailored readout circuit with active hold-off time is designed to restrain detector noise and operate the SPAD in free-running mode. We use structured illumination to reconstruct 192$\times$192 compressed images at a 4 fps frame rate. To show the superior capability of ultraviolet characteristics, we use our single-pixel imaging system to identify and distinguish different transparent objects under low-intensity irradiation, and image ultraviolet light sources. The results provide a practical solution for general ultraviolet imaging applications.
\end{abstract}

\maketitle

Ultraviolet (UV) imaging is of high value for detecting fire, surface defects, pollutants, organics, and corona discharges, which makes UV imaging of broad use in applications such as safety monitoring~\cite{zhang2019demonstration,jiang2021single}, component analysis~\cite{mcwhinney2010analysis}, medical or biological imaging~\cite{fulton1997utilizing}, remote sensing~\cite{meier1991ultraviolet,galle2003miniaturised}, and electrical maintenance. Single-photon detector (SPD) offers an improved detection efficiency, lower dark counts, and faster timing response over conventional cameras. Using SPDs in UV imaging can significantly improve imaging sensitivity and imaging distance. This can also reduce the requirement of illumination power thus avoiding radiation damage. In the field of single-photon imaging, SPD arrays are usually preferred. Unfortunately, in UV band, particularly in solar blind region, SPD arrays remain a challenging proposition and no commercial product is available at present.

\begin{figure*}[!t]\center
\resizebox{15.5cm}{!}{\includegraphics{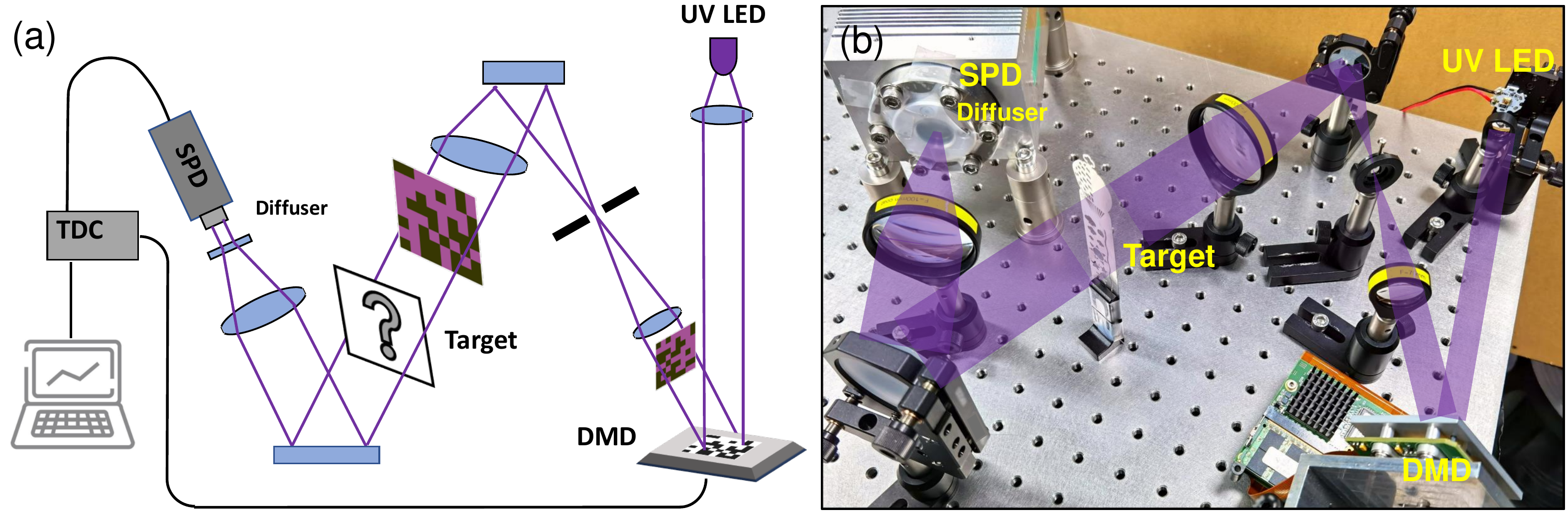}}
\caption{\textbf{Configuration of the UV single-pixel imaging system.} \textbf{(a)} Schematic of optical configuration. An UV LED operating at 310 nm emits a collimated beam, the DMD modulates the beam and projects structured patterns. The modulated beam propagates through the target, the transmitted lights are collected and detected using an UV SPD. The computer controls the DMD patterns and gets the corresponding photon-counting measurements from the TDC. \textbf{(b)} Photograph of the optical system containing the UV LED, DMD, target and SPD.}
\label{setup}
\end{figure*}

Single-pixel imaging (SPI) provides an alternative approach to realize single-photon imaging based only on a single-pixel SPD~\cite{edgar2019principles}. This method uses modulation technologies to produce patterns for either structured detection (single-pixel cameras)~\cite{duarte2008single} or structured illumination (ghost imaging)~\cite{strekalov1995observation}. Several remarkable results have emerged in the field and demonstrated the superior capability of SPI~\cite{shapiro2008computational,ferri2010differential,howland2013photon,sun20133d,zhang2015single,sun2016single,phillips2017adaptive}. Comparing with raster scanning, SPI has more efficient use of the illumination power and faster imaging speed~\cite{jiang2019scan}. Besides, SPI is capable of sensing compressively and leveraging computational techniques\cite{candes2006stable}. So far, SPI has been widely investigated at wavebands where array cameras are financially or technically difficult to access, such as near infrared~\cite{gibson2017real}, terahertz~\cite{watts2014terahertz}, and X-ray band~\cite{greenberg2014compressive}.


Recent works have demonstrated SPI in UV region based on photomultiplier tubes (PMTs)~\cite{zhang2019demonstration,jiang2021single}. However, PMTs suffer from the intrinsic issues of short lifetime, magnetic-sensitive and vacuum operation. By contrast, the photon detectors based on emergent third generation semiconductor material silicon carbide (SiC) are low-cost, small size and high stability. Among varies polytypes of SiC materials, 4H-SiC is one of the key candidates for fabricating single-photon avalanche diodes (SPADs) owing to its high wafer quality and relatively mature epitaxy technique. 4H-SiC SPAD has a wide spectral response in UV region, its peak response wavelength drops in solar blind region (260-280 nm), while its cut-off wavelength reaches 380 nm. In recent years, the performance of 4H-SiC SPADs has enhanced greatly in efficiency and noise~\cite{su2019recent,wang2017noise,dong2020after}.

In this paper, we demonstrate photon-counting SPI in UV region based on a homemade high-performance compact 4H-SiC SPD ~\cite{yu2023free}. We build a SPI setup and perform UV-imaging for both UV-absorbing substances and UV light sources. We reconstruct 192$\times$192 resolution images at a 4 fps acquisition rate using only 5000 sparse random patterns, corresponding to a compression ratio of 13.6$\%$. As an application demonstration of UV photon-counting SPI, we image different transparent objects under low-intensity UV irradiation. Given the advantages of low-cost, small size, stability and broad UV spectrum response, our system should be a useful solution for general UV imaging applications.

\begin{figure*}[!t]\center
\resizebox{15.5cm}{!}{\includegraphics{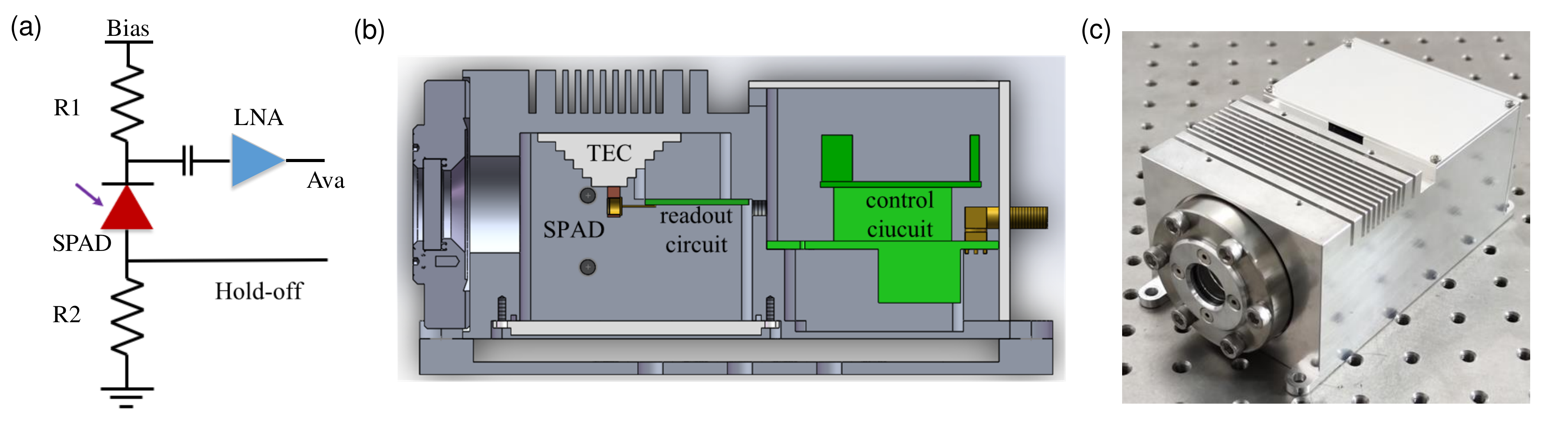}}
\caption{\textbf{Configuration of the UV SPD system.}\textbf{(a)} Readout circuit for 4H-SiC SPAD. \textbf{(b)} Structure of 4H-SiC SPD. \textbf{(c)} Photograph  of 4H-SiC SPD.}
\label{spad}
\end{figure*}

Fig.~\ref{setup} shows our UV SPI system. The UV source consists of a light-emitting diode (LED) operating at 310 nm and a 20 mm focal length UV fused silica lens. The light source emits collimated beams, and the high-speed digital micro-mirror device (DMD) modulates the beam with structured patterns. The patterns are then projected onto the imaging scene via a 4F beam-expander system. After propagate through the imaging target, the transmitted photons are detected by the UV SPD in the receiving system.

The beam-expander system has two lenses of focal lengths 7.5 mm and 20 mm. The DMD, a diaphragm, and imaging objects are placed at the Fourier plane of the two lenses, respectively. The DMD patterns are thus projected onto the target plane with 2.7$\times$ magnification. The diaphragm blocks the noisy diffracted lights from the DMD. In this experiment, the background noise is reduced from 270 kcps to 120 kcps with the diaphragm.

The resolution of the DMD chip is 1024$\times$768 and each mirror pitch has a size of 13.7 $\mu$m. Before experiments, the computer sends a sequence of patterns to the DMD's memory buffer and then controls the pattern display. In imaging process, the DMD is synchronized with UV SPD. The time-to-digital converter (TDC) counts the number of photons detected for each pattern, and the data are transferred to the computer for image reconstruction.

In the receiving system, the photons are collected using an UV fused condenser lens with 50.4 mm diameter and 100 mm focal length. A diffuser is placed in front of the UV SPD to introduce an optical blur that ensures light from the entire field-of-view can be captured by the sensitive area of UV SPD.

The SPIRAL-tap algorithm is used for image reconstruction~\cite{harmany2011spiral}. The compressed-sensing problem can be solved from the convex optimization program:
\begin{equation}
\hat{f}=\arg \min _{f ; f_{kl} \geq 0}\left\{-\log \left[\prod_{i} \operatorname{Pr}\left(Y_{i} ; f\right)\right]+\lambda \operatorname{pen}(f)\right\}
\end{equation}
The first term is the negative log-likelihood function, which considers the Poisson distribution $Y$ of photon counts. The regularization parameter $\lambda$ is optimized to obtain a suitable reconstructed image that is neither over fitted nor over smoothed.

The epitaxial structure of the 4H-SiC SPAD used in this work is grown on 4-inch n-type 4H-SiC substrate, which from bottom to top consists of a 2 $\mu$m n+ buffer layer $(N_{D} = 1 \times 10^{19}\, cm^{-3})$, a 0.5 $\mu$m p- multiplication layer $(N_{A} = 3 \times 10^{15} \, cm^{-3})$, a 0.2 $\mu$m p layer $(N_{A} = 2 \times 10^{18} \, cm^{-3})$, and a thin 0.1 $\mu$m p+ contact layer $(N_{A} = 2 \times 10^{19} \, cm^{-3})$. The fabrication process starts from mesa etching, followed by n/p contact formation and finally surface passivation. By using a photoresist reflow technique and inductively coupled plasma etching, a beveled mesa with a small slope angle of $\sim4.5^{\circ}$ is formed as a termination to suppress peak electrical field around the mesa edge. The SPAD has an effective active area of around 180 $\mu$m in diameter and is hermetically sealed in a TO-46 package for testing. The avalanche breakdown voltage of the SPAD is $\sim$178 V and the maximum avalanche gain could exceed $10^{5}$. Based on spectral response characterization, the peak responsivity of the SPAD occurs at $\sim$270 nm and reaches 0.15 A/W. At 310 nm, the responsivity is $\sim$0.046 A/W, corresponding to a unity-gain quantum efficiency of $\sim$19$\%$.

The photon arrival time in SPI application is unpredictable, so the SPD must operate in free-running mode.
Although passive quenching is a common approach for free-running SPDs, 4H-SiC SPADs suffer from a considerable afterpulse probability when using passive quenching readout circuits~\cite{wang2017noise}.
Here, we designed a passive quenching and active hold-off circuit~\cite{yu2017fully} to suppress the noise of the UV SPD. As shown in Fig.~\ref{spad}(a), R1 is a 200 k$\Omega$ passive quenching resistor, and the avalanche signal is capacitance-coupled to a low noise amplifier with around 20 dB gain. At the SPAD anode, the hold-off signal controlled by a field-programmable gate array (FPGA) is connected with a 50 $\Omega$ matched resistance R2. When an avalanche signal is detected, the hold-off signal is pulled up to 10 V after a 10 ns circuit delay, and then persists for a programmable time. The excess bias voltage in our experiment is always lower than 10 V, so SPAD exits Geiger mode during hold-off time. Since the afterpulse probability decays exponentially~\cite{dong2020after}, this method can significantly reduces the total afterpulse probability.

Fig.~\ref{spad}(b) shows the structure of the UV SPD. SPAD is installed on a thermoelectric cooler in a cooling chamber, and the readout circuit is positioned very close to the SPAD to minimize parasitic parameters. UV light illuminate the SPAD through a 19-mm-diameter observation window, and electrical signals, including bias, avalanche, hold-off and power supply, are connected to the control board through mcx connectors. The control board provides an adjustable bias voltage and hold-off signal for the SPAD, discriminates avalanche signals to 3.3 V LVTTL signals, controls the SPAD temperature to within ±0.1 $^{\circ}$C stability, and communicates with the PC software. The whole UV SPD is 160 mm$\times$100 mm$\times$75 mm in size and weight around 1.5 kg.

In experiment, the SPAD temperature is set to -40 $^{\circ}$C, bias voltage is set to 188 V and the hold-off time is 1 $\mu$s. With this parametrization, the UV SPD offers around 3$\%$ detection efficiency at 310 nm with a dark count rate of 133 kcps and negligible afterpulse probability.

\begin{figure}[t!]
\centering\includegraphics[width=8.5cm]{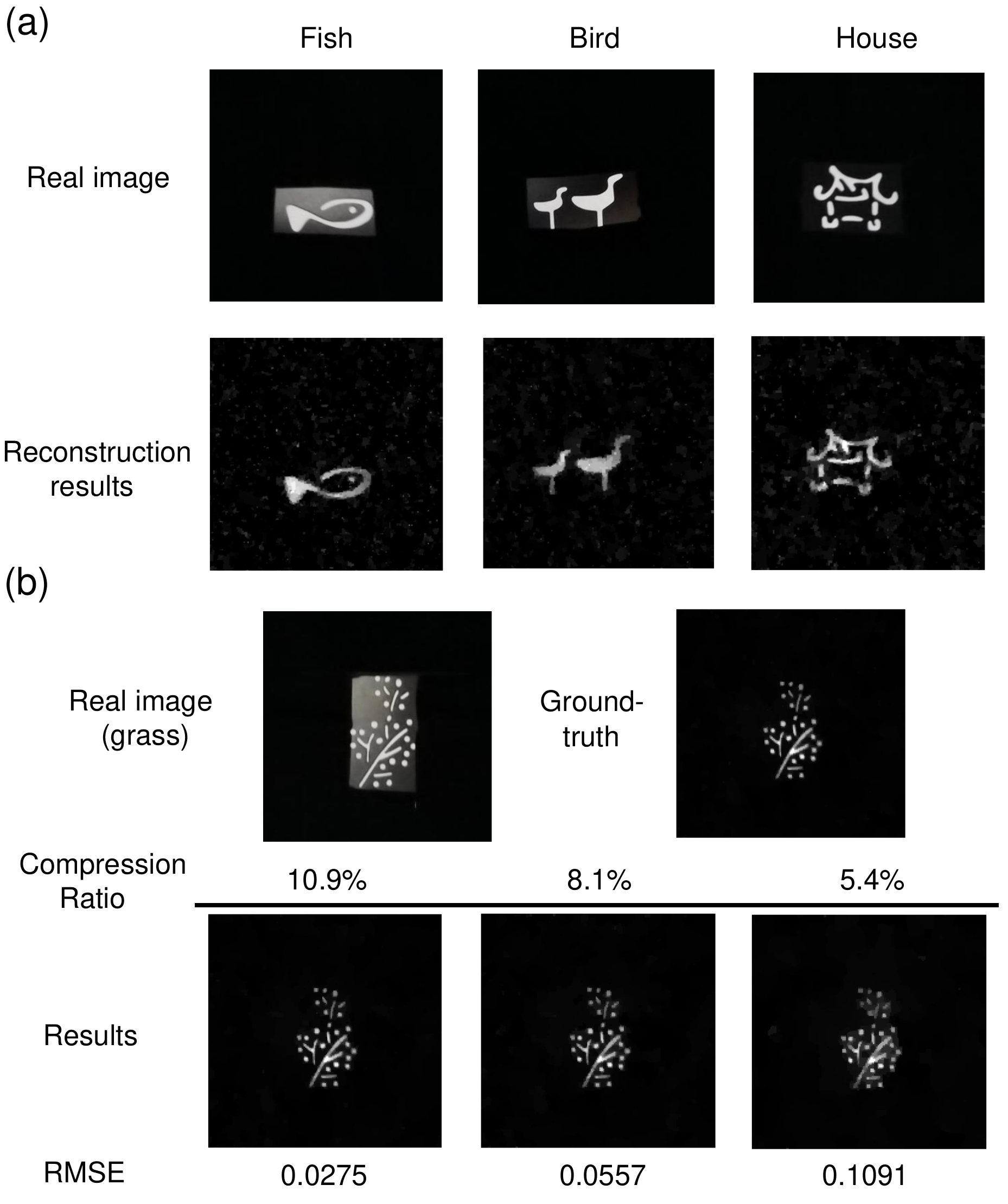}
\caption{\textbf{Imaging results of hollow designs.} \textbf{(a)} Top row shows the real images of three hollow designs: fish, bird, and house. The bottom row shows the compressed sensing imaging results with 192$\times$192 resolution, and the total data collection time for each result is 0.25 s. All three designs are well reconstructed using 5000 random binary patterns, equivalent to a compression ratio of 13.6$\%$. For each measurement pattern, the average signal photon count is about 13.\textbf{(b)} Reconstruction results with different numbers of measurements. The images are reconstructed from 2000 measurements (5.4$\%$ compression ratio) to 4000 measurements (10.9$\%$ compression ratio) are nearly identical.}
\label{result1}
\end{figure}

\begin{figure*}[t!]\center
\resizebox{16.5cm}{!}{\includegraphics{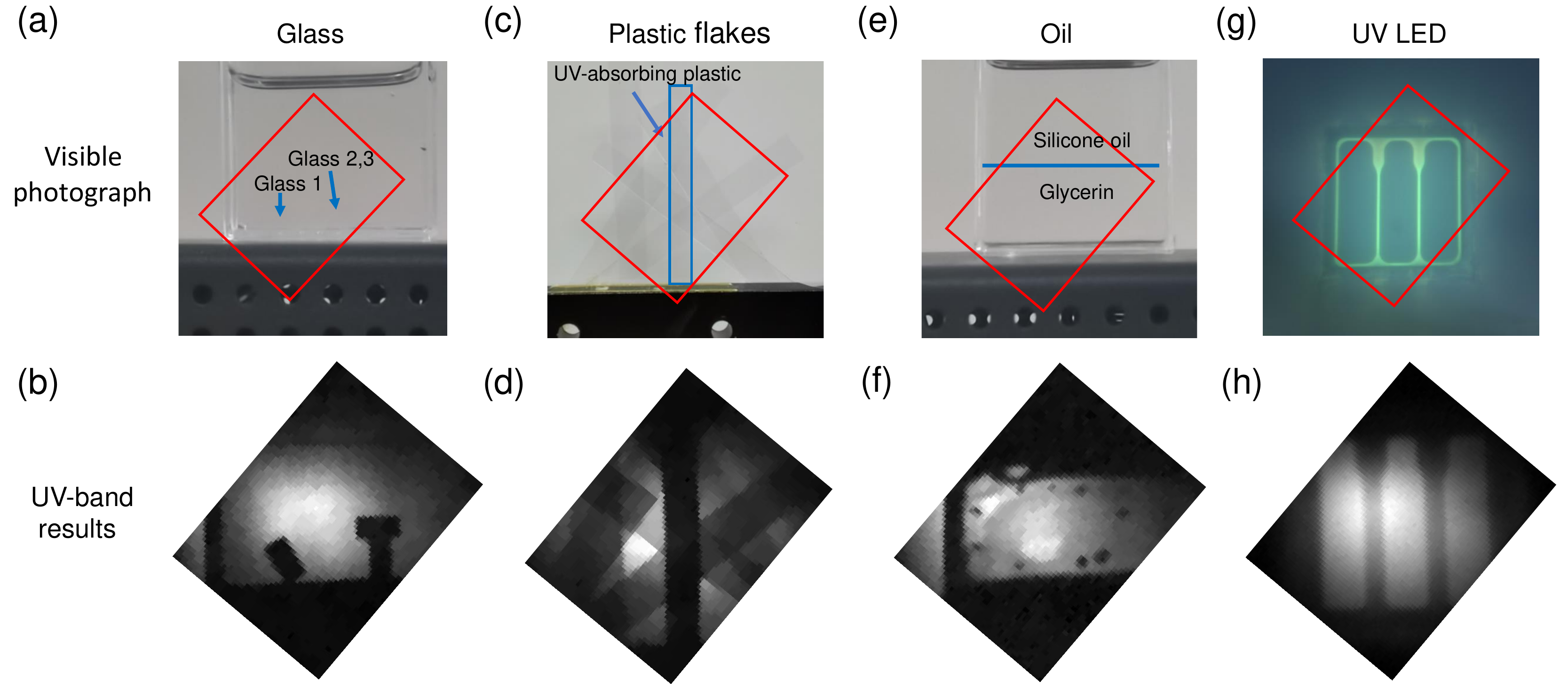}}
\caption{\textbf{UV imaging results.} To show the unique features of UV imaging, we use our system to image different transparent materials. \textbf{(a)} Three rectangular glasses in a container filled with glycerin. The glasses are not detected in the visible photograph because glycerin and glass are transparent and have similar refractive indices. \textbf{(b)} The glasses absorb in the UV, so they are clearly revealed in the glycerin. \textbf{(c)} The visible photographs show some transparent plastic flakes, one of which is made of a different material. \textbf{(d)} The UV absorption characteristics clearly reveal the unique dark plastic sheet upon UV illumination. \textbf{(e)} The container contains glycerin (bottom) and silicone oil (top), which are indistinguishable in the visible photograph. \textbf{(f)} Since silicone oil absorbs in the UV, the UV image reveals the glycerin and the residual silicone oil spots in the glycerin. \textbf{(g)} Visible photographs show only the green fluorescence of a UV LED, but does not reveal the luminous status of the LED chip. \textbf{(h)} UV imaging reveals the shape of the UV-emitting part of the LED.}
\label{result3}
\end{figure*}

\begin{figure}[!t]
\centering\includegraphics[width=6.5cm]{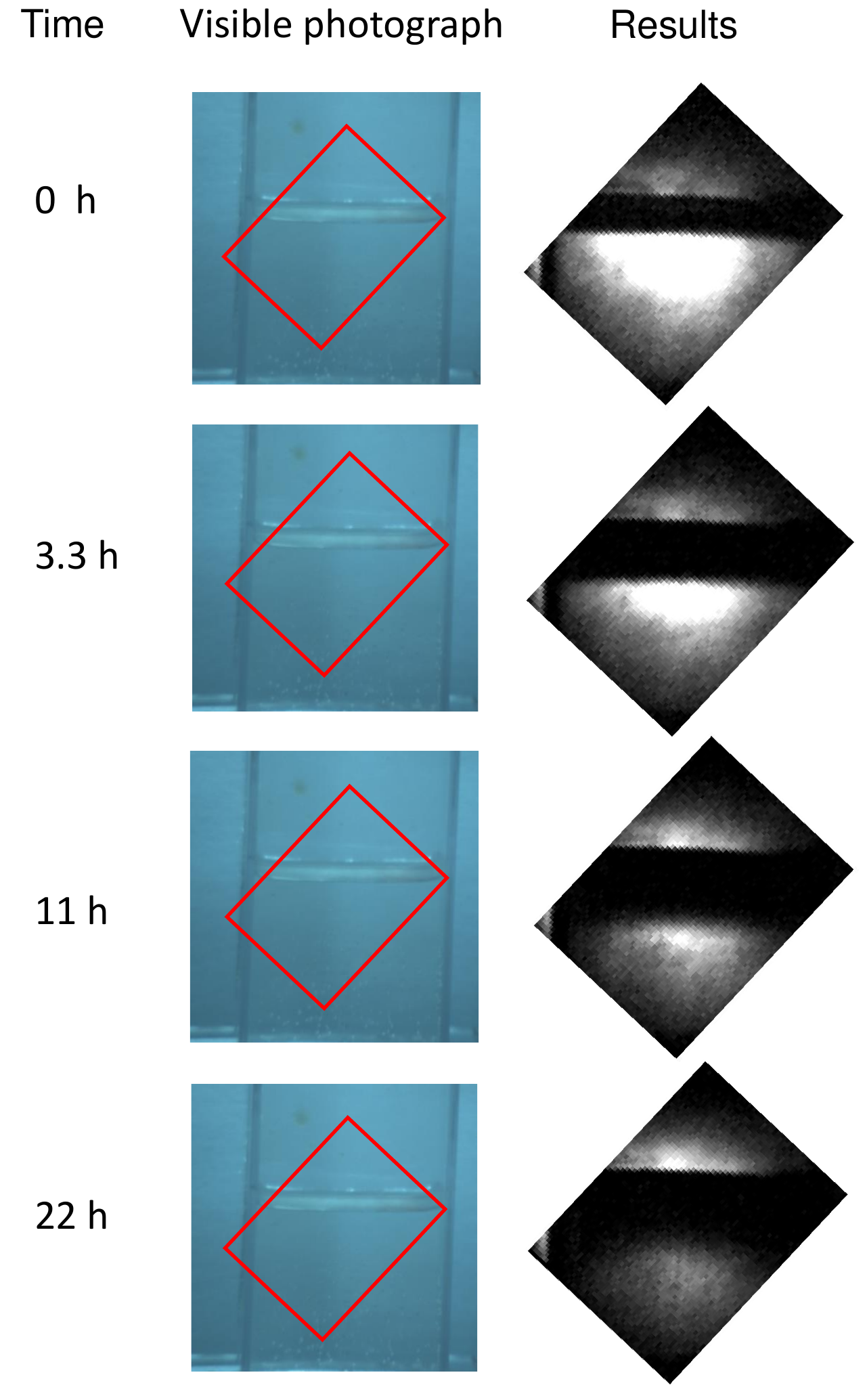}
\caption{\textbf{UV imaging results of the dissolution process.} We added some transparent juice to the water, the water is mixed with glycerin to slow the dissolution rate. The visible video shows no changes, whereas the UV imaging clearly reveals the entire dissolution process (see video in Multimedia view)}.
\label{result4}
\end{figure}


First, to demonstrate the capability of high-resolution, high-speed compressed-sensing imaging, we image three hollow iron plates with different hollow designs at a resolution of 192$\times$192, and with a data acquisition time of 0.25 s, Fig.~\ref{result1}(a) shows the results.

In experiment, due to the low intensity UV illumination and absorbance of imaging target, only about 6 pW of light reaches the photosensitive area. Thanks to the single-photon sensitivity of the UV SPD, we can still get around 256 K signal photon counts per second.

During imaging, the projected patterns of the DMD are random binary patterns with entries taking values 1 with probability 1$\%$ and 0 with probability 99$\%$.
The strategy has the following advantage: First, according to the early work~\cite{duarte2008single}, for the case of photon counting noise, imaging using sparse patterns requires fewer photons to obtain the same quality than using orthogonal basis patterns, thus making more efficient use of the detected photons. Second, the strategy provides a better signal-to-noise ratio than raster scanning and enables compressed sensing. Finally, the strategy also ensures low-intensity UV irradiation. We choose the probability value of 1$\%$ in order to guarantee a satisfied signal-to-noise ratio while avoiding the SPD saturation.

We bin 4$\times$4 pixels on the DMD as a projected pattern pixel. A total of 5000 patterns at a resolution of 192$\times$192 are pre-generated and loaded into the DMD's memory buffer. During the capture process, the illumination time of each pattern is 50 $\mu$s, and the total capture time is 0.25 s. For each measurement pattern, the average signal-photon count is about 13.

For reconstruction, we use a canonical basis and total-variation norm as a penalty, the penalty parameter is set as 0.2. Fig.~\ref{result1}(a) shows that all three designs are well reconstructed at a resolution of 192$\times$192 using compressive measurements.

We also show the reconstruction results with various numbers of measurements. We evenly extracted 4000, 3000, and 2000 measurements data from the original 5000 intensity measurements. Fig.~\ref{result1}(b) shows the recovered images of the "grass" design with varying compression ratios ranging from 10.9$\%$ to 5.4$\%$, we take the reconstruction result with 5000 measurements (13.6$\%$ compression ratio) as the groundtruth to compute the root mean square error (RMSE) of each recovered image. The results show that images reconstructed from 2000 measurements to 4000 measurements are nearly identical. Since the imaging objects are sparse and not complex, good-quality results can be reconstructed with relatively low compression ratio~\cite{candes2006stable}.

Next, we demonstrate photon-counting UV imaging to show the advantages of UV illumination. We use our system to image through substances that are transparent in the visible. We project 4096 sparse random patterns at a resolution of 64$\times$64 for imaging. The imaging results are reconstructed using compressed sensing algorithms. Figs.~\ref{result3}(a) and~\ref{result3}(b) show three rectangular glass pieces in glycerin. The glass pieces are indistinguishable in the visible photograph because the glycerin and glass are both transparent and have similar refractive indexes. In the UV, the transmittance of the glass differs from that of glycerin, so we can clearly distinguish the three pieces of glass in glycerin.

In Figs.~\ref{result3}(c) and~\ref{result3}(d), one of the plastic flakes is made of a different material, but this flake cannot be identified in the visible photographs because all the flakes are transparent. However, UV imaging clearly reveals a distinctive dark plastic flake due to its greater UV absorption. Figs.~\ref{result3}(e) and~\ref{result3}(f) show glycerin (bottom) and silicone oil (top) in a container, these are indistinguishable in the visible photograph. However, UV imaging reveals the delamination of the two liquids and the spots of silicone oil left in the glycerin.

We also perform UV imaging using single-pixel camera configurations. The imaging target is an UV LED chip, and the results appear in Figs.~\ref{result3}(g) and~\ref{result3}(h). The visible photograph shows the fluorescence light on the LED chip, whereas the UV image reveals the UV-emitting part of the LED chip.

We also imaged the dissolution process of the transparent juice in water (see Fig.~\ref{result4} and video in Multimedia view). Water is mixed with glycerin to slow the dissolution rate. Although the visible video shows no changes, the UV imaging clearly reveals the dissolution process. This result shows that the proposed system has potential applications in imaging chemical pharmaceuticals and in biological research.

In summary, this work proposes a high-performance compact SPD based on a 4H-SiC SPAD, and experimentally demonstrate SPI in UV region. The proposed imaging system can reconstruct 192$\times$192-pixel-resolution images at 4 fps acquisition rate using compressive measurements. We use this system to identify and distinguish different transparent objects. Although the targets are invisible in visible photographs, UV imaging clearly reveals the objects.
Since our system has ps-level time resolution, it can produce three-dimensional images by using a pulsed light source~\cite{sun2016single}. Besides, by further expanding the number of pixels of UV SPD to $8\times6$, we can increase the imaging resolution to $1024\times768$~\cite{herman2013higher}.
The proposed system's low-cost, small size, stability and broad UV spectrum response make it a useful solution for general UV imaging applications, especially for the photon-starved cases due to photodamage limit or light attenuation.


~\\
This work was supported by the Innovation Program for Quantum Science and Technology under Grant 2021ZD0300804, 2021ZD0303400, and 2021ZD0300300, National Natural Science Foundation of China (Grant No. 62031024), Shanghai Municipal Science and Technology Major Project (Grant No. 2019SHZDZX01), Shanghai Academic/Technology Research Leader (21XD1403800), Shanghai Science and Technology Development Funds (22JC1402900) and Key-Area Research and Development Program of Guangdong Province (2020B0303020001).

~\\
\textbf{AUTHOR DECLARATIONS}

\noindent{\textbf{Conflict of Interest}}
The authors have no conflicts to disclose.


~\\
\textbf{DATA AVAILABILITY}

The data that support the findings of this study are available from the corresponding author upon reasonable request.


\bibliography{uv}

\end{document}